\documentclass[twocolumn,prd,amsmath,superscriptaddress]{revtex4}

\usepackage{graphicx}
\usepackage{amsmath}
\usepackage{amssymb}
\usepackage{graphics}
\usepackage{color}

\usepackage{hyperref}
\usepackage{ifthen}
\usepackage{bm}

\newcommand{\Lmat}{{\mathcal L}_\mathrm{matter}}
\newcommand{\meff}{m_\phi}
\newcommand{\Veff}{V_\mathrm{eff}}
\newcommand{\rhoc}{\rho_\mathrm{c}}
\newcommand{\Pc}{P_\mathrm{c}}
\newcommand{\rstar}{{r_\star}}
\newcommand{\mstar}{{M_\star}}
\newcommand{\mlin}{{M_\mathrm{linear}}}
\newcommand{\mscr}{{M_\mathrm{screen}}}
\newcommand{\phic}{\phi_\mathrm{c}}
\newcommand{\phids}{\phi_\mathrm{dS}}
\newcommand{\phit}{{\phi_\mathrm{t}}}
\newcommand{\mt}{{m_\mathrm{t}}}
\newcommand{\Rds}{R_\mathrm{dS}}
\newcommand{\mds}{m_\mathrm{dS}}

\newcommand{\force}{{\mathcal F}}
\newcommand{\Lobs}{\Lambda_\mathrm{obs}}
\newcommand{\field}{\phi}
\newcommand{\phimin}{{\phi_\mathrm{min}}}
\newcommand{\phiscr}{{\phi_\mathrm{scr}}}
\newcommand{\rturn}{{r_\mathrm{turn}}}

\def\simlt{\lesssim}

\begin{document}

\title{The existence of relativistic stars in $\boldsymbol{f(R)}$ gravity}

\author{Amol Upadhye}
\affiliation{Kavli Institute for Cosmological Physics, Enrico Fermi Institute, University of Chicago, Chicago, IL 60637}
\author{Wayne Hu}
\affiliation{Kavli Institute for Cosmological Physics, Enrico Fermi Institute, University of Chicago, Chicago, IL 60637}
\affiliation{Department of Astronomy and Astrophysics, University of Chicago, Chicago, IL 60637}

\date{\today}


\begin{abstract}
  We refute recent claims in the literature that stars with relativistically deep potentials cannot exist in $f(R)$ gravity.  Numerical examples of stable stars, including relativistic ($G\mstar/\rstar \sim 0.1$),  constant density stars, are studied.  As a star is made larger, non-linear ``chameleon'' effects screen much of the star's mass, stabilizing gravity at the stellar center.  Furthermore, we show that the onset of this chameleon screening is unrelated to strong gravity.  At large central pressures $P>\rho/3$, $f(R)$ gravity, like general relativity, does have a maximum gravitational potential, but at a slightly smaller value:  $\left.G\mstar/\rstar\right|_\mathrm{max} = 0.345<4/9$ for constant density and one choice of parameters.   This difference is associated with negative central curvature $R$ under general
  relativity not being accessed in the $f(R)$ model, but does not apply to any known astrophysical object.  
\end{abstract}

\maketitle


\section{Introduction}

The discovery that the expansion of the universe is accelerating spurred a search for theoretical models which could explain this phenomenon.  The simplest explanation, the cosmological constant, requires extreme fine tuning in order to explain its smallness as well as its closeness to today's matter density.  This motivates the search for alternative explanations for the cosmic acceleration.  These alternatives fall into two broad classes.  In the first class, a new field, known as a ``dark energy'', comes to dominate the universe at recent times, preventing the Hubble parameter from falling as rapidly as it would in a matter-dominated universe.  The second class alters gravity itself, with modifications on large scales causing the universe to deviate from its expected deceleration today.  

Modified gravity explanations are highly constrained by our knowledge of gravity on small scales. These
hurdles include equivalence principle tests \cite{Upadhye_Gubser_Khoury_2006,Adelberger_etal_2007}, solar system measurements \cite{Will:2005va}, and the stability of gravitationally bound systems such as stars \cite{Seifert_Wald_2007,Seifert_2007}.  We consider $f(R)$ gravity, a theory in which the Ricci scalar $R$ is replaced by some function $f(R)$ in the action for gravity 
\cite{Sta80,Capozziello:2003tk,Caretal03}. 
While it has been shown that some models are consistent with solar system measurements
\cite{Hu_Sawicki_2007} and the stability of non-relativistic stars \cite{Seifert_2007}, Kobayashi and Maeda \cite{Kobayashi_Maeda_2008a} (hereafter KM) have recently claimed that 
relativistic stars are unstable in related $f(R)$ models \cite{Starobinsky_2007}.
  These arguments point to the existence of a curvature singularity in cosmologically viable theories; as the scalar field $\phi \equiv {df}/{dR} \rightarrow 1$, $R \rightarrow \infty$.   They claim that relativistic stars, with $G \mstar/\rstar \sim 0.1$, push $\phi$ right into the curvature singularity, meaning that stars which we know to exist could not in $f(R)$ gravity.  Other works, while not directly disproving this claim, have argued that the singularity may be avoided by choosing a different equation of state~\cite{Babichev_Langlois_2009}, or a model in which a divergence in the scalar field potential ensures that $\phi$ avoids the singularity~\cite{Dev:2008rx,Tsujikawa_Tamaki_Tavakol_2009}.

Here we show, through numerical computation as well as analytical argument that highly relativistic stars do indeed exist in $f(R)$ gravity.  Existence
does not hinge on a specific equation of state or choice of $f(R)$ but rather the non-linearity
of the field equations.   
The onset of non-linearity causes the field to
 stop changing with the potential via the so-called
 chameleon effect. 
Thereafter  deviations in $\phi$ from its background value are determined only by a small portion of the stellar mass and the curvature singularity is never reached in a static star. 
Non-linearity in the equations of motion make the numerical solutions difficult to attain, which has obscured these points in the literature.
Nonetheless, we have numerically confirmed the existence of ultra-relativistic stars with potentials $G \mstar/\rstar$ up to $0.345$ and central pressures much greater than their energy densities.

Furthermore, we show that the onset of non-linear chameleon effects has nothing to do with strong gravity.  They will generically arise when the
gravitational potential $G \mstar/\rstar$ exceeds the field distance between the background
value and the curvature singularity, which depends on the $f(R)$ function itself.   In KM
\cite{Kobayashi_Maeda_2008a}, this distance was taken to be of order $0.1$.   In fact
this distance must be $\simlt 10^{-6}-10^{-5}$ to remain compatible with local tests of
gravity due to the finite extent of our galaxy \cite{Hu_Sawicki_2007}.  

The paper is organized as follows.  After introducing $f(R)$ theory and its application to stars in Sec.~\ref{sec:background}, we present our numerical solutions in Sec.~\ref{sec:numerical_solutions}, including relativistic as well as non-relativistic stars.  Sec.~\ref{sec:screening} employs analytic arguments in the linear
and non-linear regimes to elucidate how chameleon screening allows the field to avoid the curvature singularity.  We conclude in Sec.~\ref{sec:conclusion}.


\section{Formalism}

We briefly review the equations governing  $f(R)$ theory in Sec.~\ref{subsec:theory}, apply
them to stellar, static, spherically symmetric cases in Sec.~\ref{subsec:stars}, and
specialize to the Starobinsky $f(R)$ model  \cite{Starobinsky_2007} in Sec.~\ref{subsec:starobinsky}.

\label{sec:background}

\subsection{${f(R)}$ theory}
\label{subsec:theory}

Replacing the Ricci scalar $R$ in the Einstein-Hilbert action defining general relativity by a function $f(R)$ results in the action
\begin{equation}
S = \int d^4x \sqrt{-g} \left( \frac{f(R)}{16\pi G} + \Lmat \right)
\label{e:SfR}
\end{equation}
and the modified Einstein equation
\begin{equation}
\field R_{\mu\nu} - \nabla_\mu \nabla_\nu \field + g_{\mu\nu}\Box \field - \frac{1}{2} f g_{\mu\nu} 
= 8\pi G T_{\mu\nu}.
\label{e:mod_ein}
\end{equation}
The quantity $\field \equiv {df}/{dR}$ behaves as a scalar field coupled to matter and the metric, as can be seen by taking the trace of Eq.~(\ref{e:mod_ein}),
\begin{eqnarray}
\Box \phi 
&=& \frac{8\pi G}{3} T 
+ \frac{1}{3}\left[ 2 f(R(\phi)) - \phi R(\phi)\right] \nonumber\\
&\equiv& \frac{8\pi G}{3} T + \frac{dV}{d\phi}
\equiv \frac{\partial \Veff}{\partial \phi},
\label{e:eomphi}
\end{eqnarray}
where  $R$ is now an implicit function of $\phi$.  Evidently the scalar is a chameleon field
\cite{Gubser_Khoury_2004,Khoury_Weltman_2004,Brax_etal_2004}; its self interaction $V(\phi)$ and its coupling to matter give the field a constant value in a medium of constant $T = T^\mu_{\hphantom{\mu}\mu} = -\rho+3P$ which
also determines its mass.  We will see that non-linear effects associated with changes
in $T$ between two different media are crucial for understanding $f(R)$ solutions in stars.

\subsection{Spherical stars}
\label{subsec:stars}

Since we intend to study stars, we assume a spherically symmetric metric,
\begin{equation}
ds^2 = -N(r) dt^2 + \frac{dr^2}{B(r)} 
+ r^2(d\theta^2 + \sin^2\theta d\varphi^2).
\label{e:metric}
\end{equation}
With this metric, the field equation (\ref{e:eomphi}) for static solutions becomes
\begin{equation}
\left[\phi'' + \left(\frac{2}{r} + \frac{N'}{2N} + \frac{B'}{2B}\right)\phi'
\right] B
=
\frac{dV}{d\phi}
- \frac{8\pi G}{3}(\rho-3P) .
\label{e:sphi}
\end{equation}
The system is completed by the $(tt)$ and $(rr)$ components of the modified Einstein equations,
\begin{eqnarray}
\frac{\phi}{r^2}(-1 + B &+& rB')
+ \left[\phi'' + \left(\frac{2}{r} + \frac{B'}{2B}\right)\phi'\right]B 
\nonumber\\
&=&
-8\pi G \rho - \frac{1}{2}\phi R(\phi) + \frac{1}{2} f(R(\phi)),
\label{e:sB}
\\
\frac{\phi}{r^2}\bigg(-1 + B &+& \frac{rBN'}{N}\bigg)
+ \left(\frac{2}{r} + \frac{N'}{2N}\right) \phi' B 
\nonumber\\
&=&
8\pi G P - \frac{1}{2}\phi R(\phi) + \frac{1}{2} f(R(\phi)) ,
\label{e:sN}
\end{eqnarray}
the equation of hydrostatic equilibrium,
\begin{equation}
P' = -\frac{N'}{2N}(\rho + P)\, ,
\label{e:sP}
\end{equation}
and equation of state $\rho = \rho(P)$ for the matter.  We follow KM \cite{Kobayashi_Maeda_2008a} and assume a constant density, $\rho(P) = \rhoc$,
and central pressure $P_c$, but this may easily be generalized.
Here, and throughout the paper, primes denote derivatives with respect to $r$.

Boundary conditions for $\phi'$, $P$, $N$, and $B$ can be specified at the center.  Continuity of the gradient of $\phi$ at the center of the star requires $\phi'(0)$.  $P(0)$ is set to a specified central pressure $\Pc$.  In order to facilitate comparison with KM, we take $N(0) = B(0) = 1$, amounting to a rescaling of the time coordinate.  

The remaining boundary condition for the field is more complicated. 
We again follow KM and take the exterior of the star to be empty save for the $\phi$ field and any effective cosmological
constant that its value implies.  In general relativity, the exterior metric would be the
 Schwarzschild-de Sitter spacetime, which has a horizon $r_\mathrm{N}=r_\mathrm{B}\approx \sqrt{3/\Lambda}$  where $N(r_\mathrm{N})=0$ and $B(r_\mathrm{B})=0$.  In $f(R)$ gravity, $N$ and $B$ do not necessarily vanish at the same position.  
 However, the curvature invariant
  $R_{\alpha\beta\gamma\delta}R^{\alpha\beta\gamma\delta}$ diverges at $r_\mathrm{N}$ if $r_\mathrm{N} \neq r_\mathrm{B}$.  In order to ensure that a solution of the equations of motion yields a well-behaved cosmology at the horizon, our final boundary condition must be $r_\mathrm{N} = r_\mathrm{B}$. 
    The field equation (\ref{e:sphi}) and linearity in $N$ and $B$ around $r_\mathrm{N}$ indicates that this boundary condition is equivalent
    to taking  $B' \phi'  -  V_{,\phi} |_{r\rightarrow r_\mathrm{N}}=0$.

For a star with specified $\rhoc$ and $\Pc$, we find $\phi(r)$, $N(r)$, $B(r)$,  and $P(r)$ using a shooting method.  We begin by guessing a central value for the field, $\phi(0) = \phic$.   With
the boundary and normalization conditions above, we can solve the above system of equations, thereby finding $B'\phi'-V_{,\phi}$ at the horizon $r_\mathrm{N}$.  Suppose we have two guesses, $\phi_{\mathrm{c},-}$ and $\phi_{\mathrm{c},+}$, for which $B'\phi'-V_{,\phi}<0$ and $>0$, respectively.  Since the system of equations (\ref{e:sphi}-\ref{e:sP}) provides a continuous mapping from $\phic$ to $B'\phi'-V_{,\phi}$ at the horizon, there must be a value $\phic$ between $\phi_{\mathrm{c},-}$ and $\phi_{\mathrm{c},+}$ for which the boundary condition at the horizon is satisfied.  We find an 
improved estimate of the true solution by iteratively bisecting this interval until the dimensionless boundary condition, $b(\phic) \equiv r_\mathrm{N}(\phi'-V_{,\phi}/B')$, evaluated when $N$ drops to $10^{-15}$, is within $10^{-5}$ of zero.
  Note that for stars that are much smaller than the horizon, 
$\phi \approx \phids$ and nearly constant, so $\phi'$ and $V_{,\phi}$ both vanish at $r_\mathrm{N}$.

Even in cases where it is numerically difficult to find the solution to the required
precision, a solution should still exist if both under and overshoot solutions also exist.
This is simply the overshoot-undershoot argument of \cite{Coleman_1977,Coleman_deLuccia_1980}, applied to spherical stars in an $f(R)$ analogue of a Schwarzschild-de Sitter background, rather than to spherical bubbles in a de Sitter background.  Similarly to those references, our field equation (\ref{e:sphi}) looks like the equation of motion of a particle at ``position'' $\phi$, as a function of ``time'' $r$, in a potential $U(\phi)\equiv -V(\phi)$, with a time-dependent friction term $\sim 2/r$, driven by a time-dependent force $\force = 8\pi G T / 3$.  For stars much smaller than the horizon, the 
vanishing of $dV/d\phi$ at the horizon corresponds to the field stopping at 
the maximum of $U(\phi)$, corresponding to the false vacuum in \cite{Coleman_deLuccia_1980}.

\subsection{Starobinsky $f(R)$ model}
\label{subsec:starobinsky}

The above discussion applies to all $f(R)$ models.  In order to proceed, we must specify a model.  To test the KM claim, we follow them in choosing the model of Starobinsky \cite{Starobinsky_2007},
\begin{equation}
f(R) = R + \lambda R_0 \left[\left(1 + \frac{R^2}{R_0^2}\right)^{-n} - 1\right],
\label{e:staro_fR}
\end{equation}
where $n$, $\lambda$, and $R_0$ are free parameters, and the field 
\begin{equation}
\phi 
= 
1 - 2n\lambda \frac{R}{R_0} \left( 1 + \frac{R^2}{R_0^2}\right)^{-(n+1)}.
\label{e:staro_phi}
\end{equation}

\begin{figure}[tb]
\begin{center}
\includegraphics[angle=270,width=3.5in]{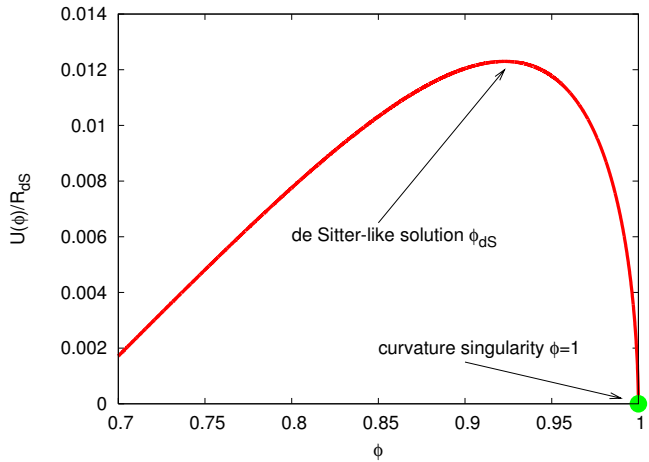}
\caption{Inverted potential $U(\phi) = -V(\phi)$, up to an arbitrary additive constant, for $n=1$ and $x_1 = 3.6$. \label{f:U}}
\end{center}
\end{figure}
\begin{figure}[tb]
\begin{center}
\includegraphics[angle=270,width=3.5in]{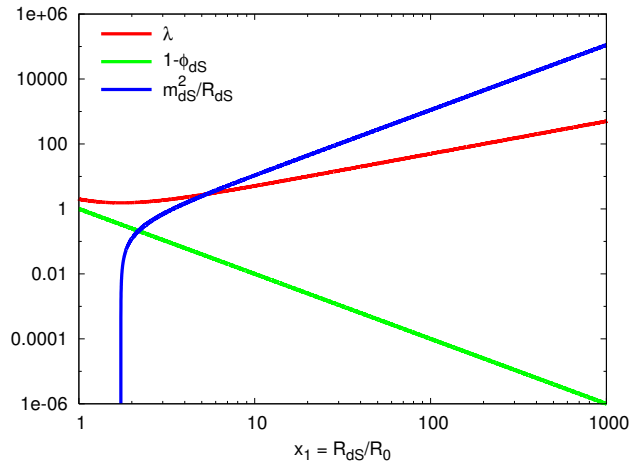}
\caption{Properties of the potential $U(\phi)$ as a function of $x_1 \equiv \Rds/R_0$ for $n=1$.  Increasing $x_1$---that is, decreasing $R_0$ at fixed $\Rds = 4\Lobs$---pushes the maximum of the potential closer to the curvature singularity. Note that the effective mass $\mds$ vanishes as $x_1\rightarrow \sqrt{3}$.  \label{f:dS}}
\end{center}
\end{figure}

The potential $U(\phi)$, defined up to an additive constant by 
\begin{eqnarray}
\frac{dU}{d\phi} 
&=& 
-\frac{dV}{d\phi} 
=
\frac{1}{3}(\phi R - 2f) 
\nonumber\\
&=& 
-\frac{1}{3}R 
+ \frac{2\lambda R_0}{3}
\left[1 - \frac{1+(n+1)R^2/R_0^2}{(1+R^2/R_0^2)^{n+1}} \right],
\end{eqnarray}
is shown in Fig.~\ref{f:U}.  Note that $\phi=1$ corresponds to a curvature singularity, $R = \infty$.  This is a common feature of $f(R)$ where the modification to the Einstein-Hilbert action
vanishes at high curvature, including Starobinsky models with $n>0$ and the broken power law models of Hu and Sawicki \cite{Hu_Sawicki_2007}.  For $\phi$ near the singularity, $R \gg R_0$, so $U'(\phi) \approx -R(\phi)/3$.  Eq.~(\ref{e:staro_phi}) implies $(R/R_0)^{2n+1} \approx 2n\lambda/(1-\phi)$, so 
\begin{equation}
\left.\frac{dU}{d\phi}\right|_{R \gg R_0}
\approx
-\frac{1}{3} R_0 \left(\frac{2n\lambda}{1-\phi}\right)^{\frac{1}{2n+1}}.
\label{e:dU1}
\end{equation}
The effective chameleon mass $\meff^2 \equiv -U''(\phi)$ near the singularity is 
\begin{eqnarray}
\left.\meff^2\right|_{R \gg R_0}
&=&
\frac{1}{3}\left(\frac{\phi}{d\phi /dR} - R\right)
\approx
\frac{R_0}{6n(2n+1)\lambda} \left(\frac{R}{R_0}\right)^{2n+2}
\nonumber\\
&=&
\frac{R_0}{6n(2n+1)\lambda}
\left(\frac{2n\lambda}{1-\phi}\right)^\frac{2n+2}{2n+1}.
\label{e:meff}
\end{eqnarray}
As the singularity is approached, the effective mass and the slope of the potential diverge.

Also, $U$ has a maximum at some field value $\phids$. Far from a star, $\phi$ will approach $\phids$, corresponding to a background, de Sitter-like spacetime.  In order to make the de Sitter background universe resemble ours, we require that $\Rds \equiv R(\phids) = 4\Lobs$, where $\Lobs$ is the observed value of the cosmological constant.  With this constraint, the two constants $\lambda$ and $R_0$, as well as the position of the maximum $\phids$, are specified by the parameter $x_1 \equiv \Rds/R_0$, as shown in Fig.~\ref{f:dS}.  
The figure also shows the effective chameleon mass at $\phids$, $\mds^2 \equiv -U''(\phids)$.

Note that KM chose a value of $x_1 = {\cal O}(1)$ for their tests and hence a value
of $1-\phids \sim 0.1$.  We shall see that that choice is responsible for the appearance of
non-linear effects only in relativistic stars, but is not viable due to solar system
tests of gravity \cite{Hu_Sawicki_2007}.

\section{Numerical Solutions}
\label{sec:numerical_solutions}

\begin{figure*}[tb]
\begin{center}
\includegraphics[angle=270,width=3.5in]{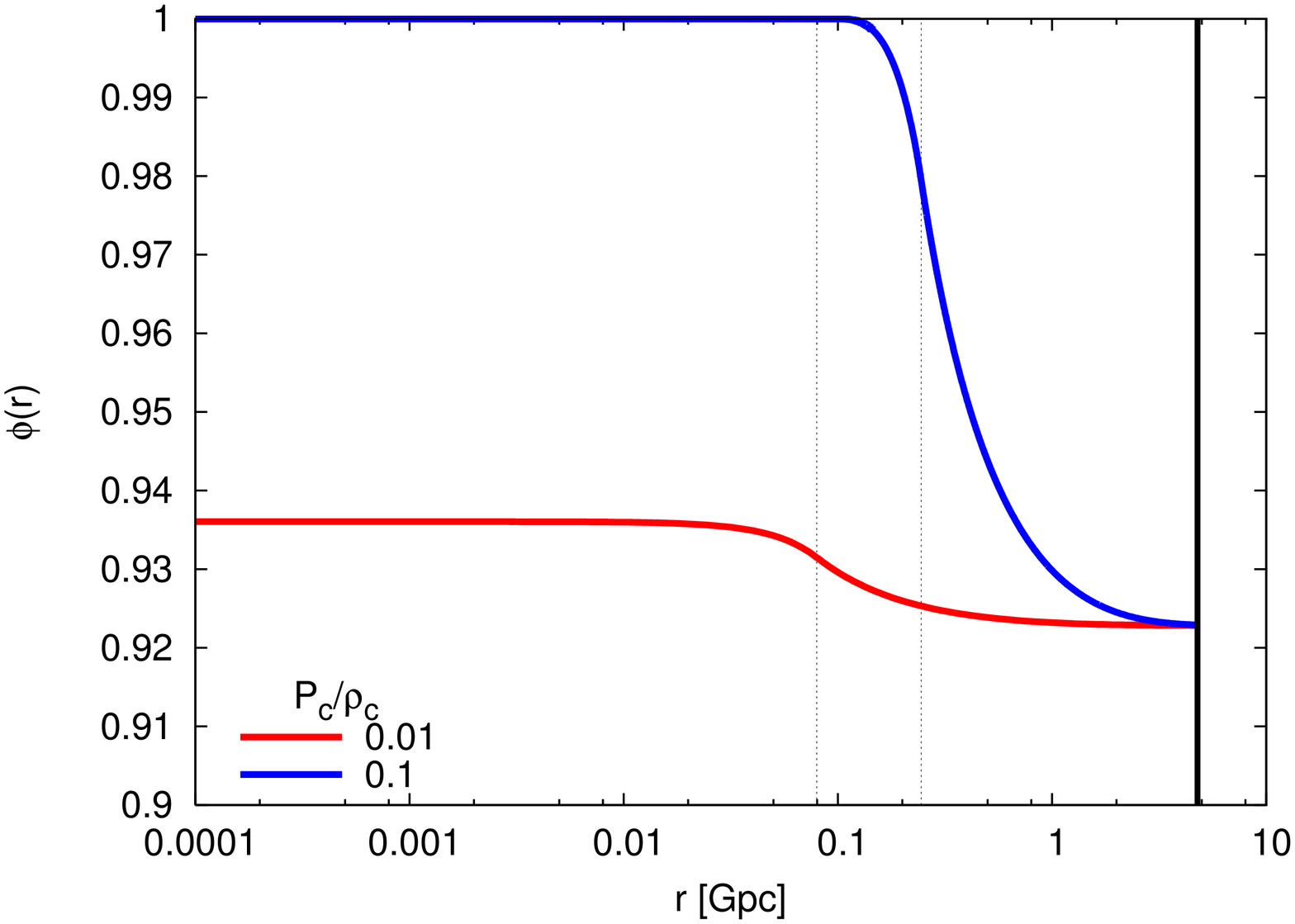}
\includegraphics[angle=270,width=3.5in]{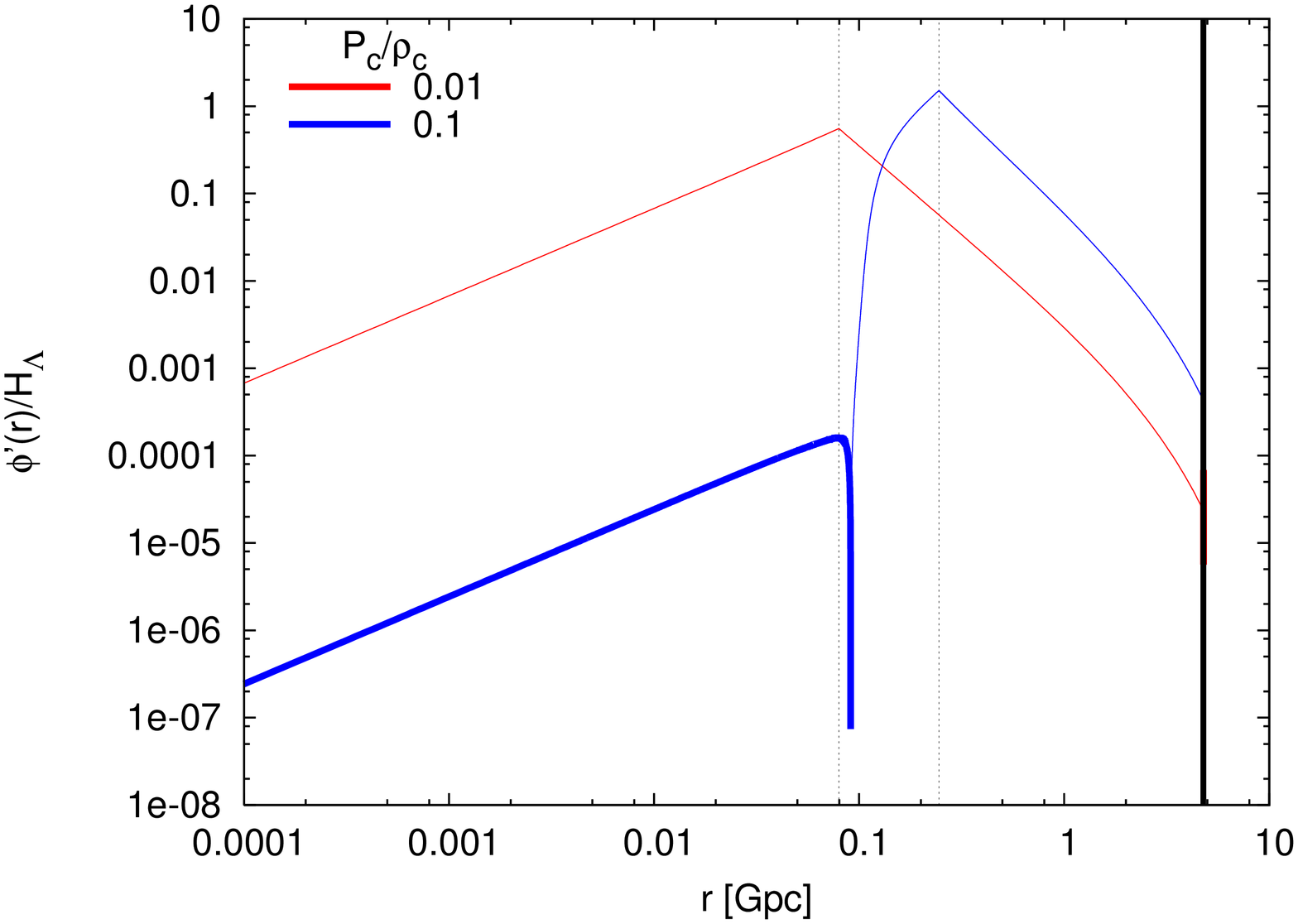}
\caption{Chameleon field $\phi$ (left) and its derivative $\phi'$ (right) in a non-relativistic star with $\Pc/\rhoc=0.01$ and $G\mstar/\rstar=0.0139$, as well as a relativistic star with $\Pc/\rhoc=0.1$ and $G\mstar/\rstar=0.131$. We have assumed $n=1$, $x_1=3.6$, and $\rhoc = 100 \rho_\Lambda$.  The solid vertical line denotes the position of the horizon, and the dotted lines are the stellar radii, with the larger radius corresponding to the relativistic star.  In the plot of $\phi'$, on the right, thick lines represent positive $\phi'$ and thin lines represent negative $\phi'$.  $H_\Lambda \equiv \sqrt{\Lambda/3} \approx 1/r_\mathrm{N}$ is the horizon scale in the analogous general relativistic spacetime.  \label{f:phi}}
\end{center}
\end{figure*}

We discuss numerical considerations in Sec.~\ref{subsec:considerations} that lead to the
choice of stellar parameters for which we give numerical solutions in Sec.~\ref{subsec:examples}.   These examples are chosen to have gravitational potentials
that are comparable to or exceed astrophysical stars albeit at a much lower density.
These considerations and those of the next section suggest that stars with realistic densities should also
exist, even though these cases are numerically intractable with our techniques. 

\subsection{Numerical considerations}
\label{subsec:considerations}

A chameleon field can change on distance scales of order its Compton wavelength $\meff^{-1}$ (see Eq.~\ref{e:meff}).     Because Yukawa-like error modes grow as
$e^{\meff r}/r$, numerical solutions become rapidly intractable as this scale
becomes much smaller than the computational domain.

 In order to study the chameleon field numerically in a star of radius $\rstar$ in a brute force implementation, we must use more than $\meff r_{\rm max}$ integration steps.   For our calculation $r_{\rm max}=r_N$, the horizon, but in general
$r_{\rm max} \gg \rstar$ in order to match an exterior solution.
  For small stars, the chameleon field is in the linear regime, where $\phi$ is only slightly perturbed from $\phids$, and $\meff(\phi) \approx \mds$ \cite{Chiba_2003,Erickcek_2006}.  Fig.~\ref{f:dS} shows that $\mds$ lies within a few orders of magnitude of the horizon scale $H_\Lambda \equiv \sqrt{\Lambda/3}$ for a large range of $x_1$. Because $H_\Lambda \rstar \ll 1$, the chameleon equations are numerically tractable for small stars.  

On the other hand, for large stars $\phi$ will be in the non-linear regime, $R\gg R_0$.  Assuming $R \approx 8\pi G(\rhoc-3\Pc)$, for which $\phic$ minimizes the effective potential $\Veff$ at the center of the star, we have 
\begin{eqnarray}
\meff 
&\approx& 
\left(\frac{R_0}{6n(2n+1)\lambda}\right)^{1/2} 
\left(\frac{R}{R_0}\right)^{n+1}
\nonumber\\
\Rightarrow 
\meff \rstar 
&\sim& 
R_0^{1/2} 
\left(\frac{8\pi G \rhoc}{R_0}\right)^{n+1}
\left(\frac{12 \Pc}{8\pi G \rhoc^2}\right)^{1/2}
\nonumber\\
&\sim&
\left(\frac{\Pc}{\rhoc}\right)^{1/2}
\left(\frac{8\pi G \rhoc}{R_0}\right)^{n+1/2},
\label{e:nsteps}
\end{eqnarray}
where the non-relativistic approximations $\Pc \ll \rhoc$ and $\rstar^2 = 12 \Pc / (8\pi G \rhoc^2)$ have been used.  For $x_1$ of order unity, $R_0 \sim 8\pi G \rho_\Lambda$.  Thus, if we want $\meff \rstar \lesssim 1000$, then we must have $\rhoc \lesssim 100 \rho_\Lambda$ for $n=1$, and even lower $\rhoc$ for larger $n$.

On top of these issues, the shooting technique exacerbates the difficulty in finding solutions
that satisfy the exterior boundary condition.   Again because of the Yukawa-like 
nature of the solutions,  the central field value $\phi_c$ must be exponentially tuned
to give the correct boundary value (see Sec.~\ref{subsec:screened_stars} for a more extended
discussion).   
For example, consider the star with $n=1$, $x_1=3.6$, $\rhoc=100\rho_\Lambda$, and $\Pc/\rhoc = 0.1$, with the boundary condition $b(\phic) \equiv r_\mathrm{N}( \phi'-V_{,\phi}/B') = 0$ at the horizon.  We find numerically that, if we want $|b(\phic)|<10^{-5}$ at the horizon, then $\phic$ must be tuned to within $1.5\times 10^{-37}$ of its correct value.  Requiring $|b(\phic)|<10^{-10}$ means tuning $\phic$ to within $1.5\times 10^{-42}$ of its correct value.
To avoid these issues, relaxation
methods can be applied instead  \cite{Hu_Sawicki_2007} but we choose
a shooting method to test the KM claim directly.  

KM  note that denser stars are more numerically difficult, and argues that only the gravitational potential is relevant as a measure of the star's size; an instability in a large, low-density star should persist in a smaller, denser star of the same gravitational potential.  KM then go on to choose $\rhoc = 2\times 10^6 \rho_\Lambda$, far from the density of a typical
star. However, even this density is too large.  By Eq.~(\ref{e:nsteps}), integration of the equations of motion for a star with this density would require billions of integration steps
for a brute force approach, and shooting compounds this problem by requiring a large
number of these solutions to iterate to the proper boundary conditions.


This numerical difficulty appears as the field fluctuations transition to the non-linear regime 
where the Compton wavelength shrinks substantially in the stellar interior.  If we choose a large, constant density, and gradually increase the ``size'' $G\mstar/\rstar$ of the star, then the integration of the equations of motion will rapidly become more difficult as the non-linear regime is approached.  In the case of stars needing billions of integration steps, truncation errors may make integration impossible.   Nonetheless this numerical difficulty does not
imply that solutions do not exist. 

\subsection{Example Solutions}
\label{subsec:examples}

Contrary to the claim of KM \cite{Kobayashi_Maeda_2008a}, we find that stellar solutions
exist at large gravitational potentials, $G\mstar/\rstar > 0.1$.  As discussed above, we set $n=1$ and $\rhoc = 100 \rho_\Lambda$ in order to keep the problem tractable.  We also set $x_1 = 3.6$, unless otherwise specified, for ease of comparison with KM.  In each case we have integrated the equations of motion (\ref{e:sphi}-\ref{e:sP}) directly, using a Runge-Kutta-Fehlberg mixed 4th/5th order algorithm with variable step sizes.  We also employ the arbitrary precision arithmetic package CLN~\cite{CLN}, and all our computations use at least $50$ decimal places.  Fig.~\ref{f:phi} shows $\phi(r)$ and $\phi'(r)$ for a non-relativistic star with $\Pc=10^{-2}\rhoc$ and $G\mstar/\rstar = 0.0139$, as well as a relativistic star with $\Pc=10^{-1}\rhoc$ and $G\mstar/\rstar = 0.131$.  

For comparison purposes, we have also attempted to extend our computations to higher a central density $\rhoc = 2\times 10^6 \rho_\Lambda$, as in KM.  For unsaturated stars with pressures of $10^{-4}\rhoc$ and $0.05\rhoc$, we find solutions that agree with those in KM.  Stars in the non-linear regime are numerically intractable, as expected.  For $\Pc = 0.1\rhoc$,  numerical instabilities prevent us from following $\phi(r)$ beyond $r \sim 10^{-5}\rstar$.  Our $\phi'(r)$ oscillates with an increasing amplitude about some central value, and eventually flies off to large positive or negative values, leading to an undershoot or an overshoot, respectively.  


\begin{figure*}[tb]
\begin{center}
\includegraphics[angle=270, width=3.5in]{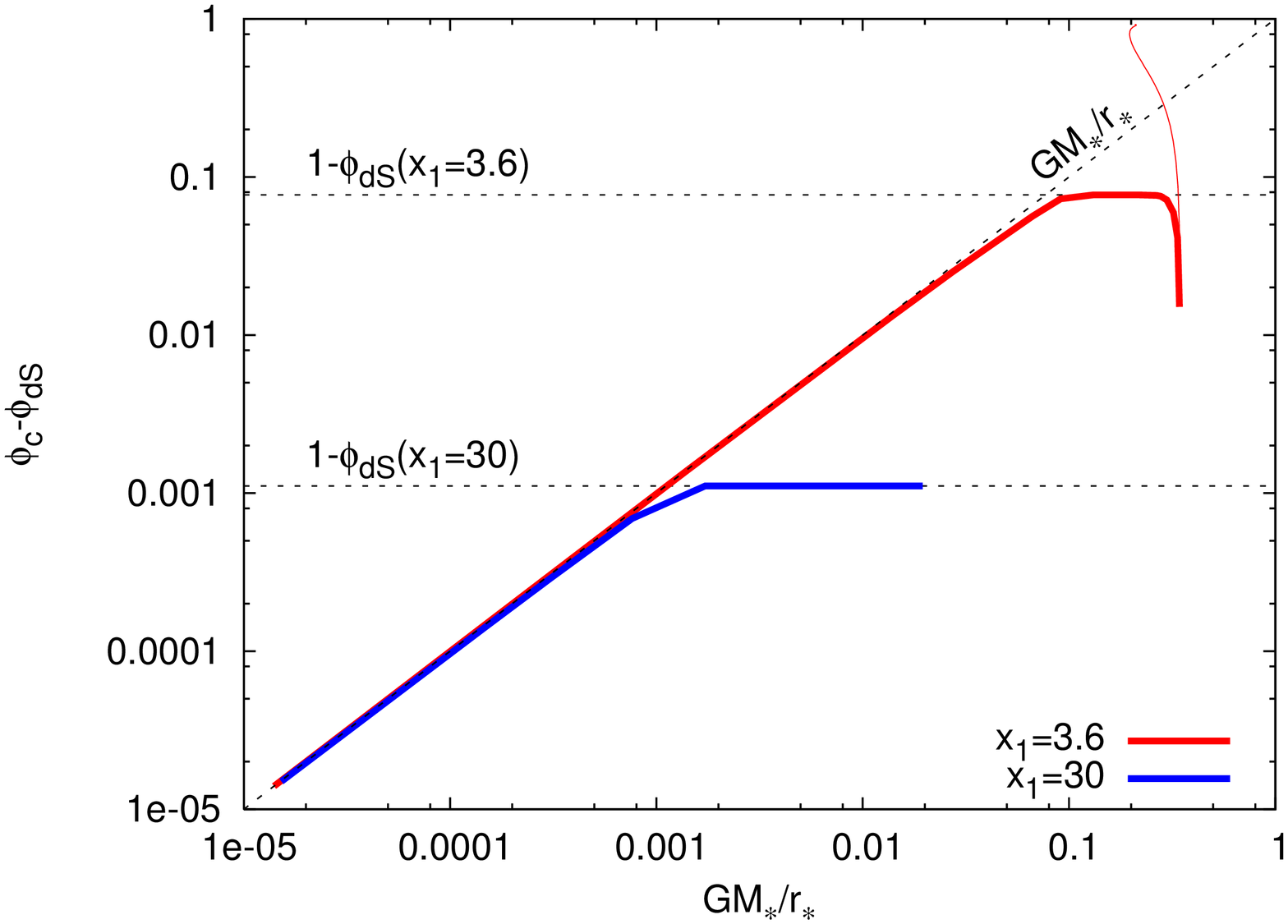}
\includegraphics[angle=270, width=3.5in]{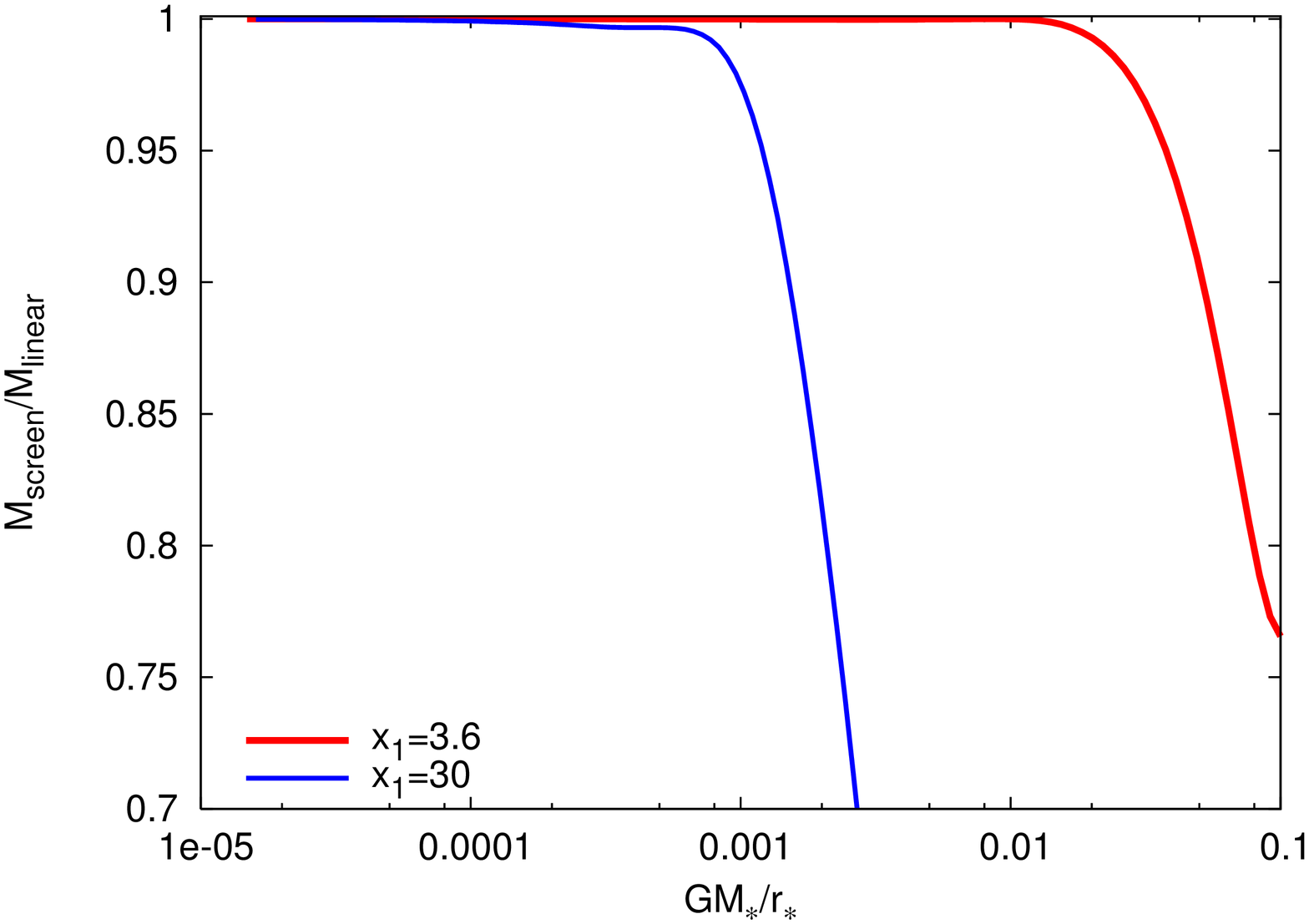}
\caption{Chameleon screening for two values of $x_1$, with $n=1$ and $\rhoc = 100 \rho_\Lambda$.  Left: The chameleon field's deviation $\phic-\phids$ from its background value rises with gravitational potential until $G\mstar/\rstar \approx 1-\phids$, after which the chameleon becomes significantly less responsive to further increases in potential.  The thin curve denotes negative values of $\phic-\phids$, reached at very high pressures, $\Pc \gtrsim \rhoc$.  Right: As the chameleon enters the non-linear regime, its source $\mscr$ decreases with respect to its linear regime source $\mlin$.  This screening becomes important around $G\mstar/\rstar \approx 1-\phids$. \label{f:screen}}
\end{center}
\end{figure*}

Our lower central density choice makes numerical solutions tractable for the full
range of central pressures.
It is evident from Fig.~\ref{f:phi}~(left) that increasing $P_c$ from $10^{-2}\rhoc$ to $10^{-1}\rhoc$ (with corresponding increases in the gravitational potential) causes a greater perturbation in the field $\phic$ from its background value toward the curvature singularity.

The mechanical analogy  between $\phi$ and a particle in the potential $U(\phi)$ is helpful
in understanding the behavior of the field and why it does not hit the curvature singularity
as $P_c$ is raised further.
 From the shape of the potential shown in Fig.~\ref{f:U}, we see that larger $\phic$ corresponds to a larger magnitude of the slope $U_{,\phi}$ of the potential, and Eq.~(\ref{e:dU1}) implies that this slope diverges as $\phic \rightarrow 1$.  On the other hand, a larger pressure corresponds to a smaller value of the ``force'' term $|{\mathcal F}| = 8\pi G(\rho-3P)/3$ in the field equation of motion (\ref{e:sphi}).   $\mathcal F$ pushes the field to lower values, allowing it to roll towards the peak $\phids$ of the potential. 
 
 Since increasing $\Pc$ causes the slope $|U_{,\phi}|$ to rise and the force $|\mathcal F|$ to fall, there must be some central pressure beyond which the slope of the potential overwhelms the force, and the field cannot decrease near the center.  From Eq.~(\ref{e:sphi}) we see that the threshold value $\phit$ at which the force at $r=0$ precisely cancels the slope of the potential is the minimum of the effective potential $\Veff$,
\begin{eqnarray}
0 
&=&
\frac{\partial \Veff}{\partial \phi} 
=
-\left.\frac{dU}{d\phi}\right|_{\phit} 
- \frac{8\pi G}{3}(\rhoc-3\Pc)
\nonumber\\
\Rightarrow
\phit
&\approx&
 1 - 2n\lambda \left[\frac{R_0}{8\pi G (\rhoc - 3\Pc)}\right]^{2n+1}.
\label{e:phit}
\end{eqnarray}

One argument against the existence of relativistic stars is that it is possible to increase $\Pc$ until $\phic > \phit$.  In such a star, $\phi$ will increase with $r$ near the stellar center, approaching the singularity $\phi=1$.  However, the curve corresponding to $\Pc/\rhoc=0.1$ in Fig.~\ref{f:phi}~(right) shows that it is possible for $\phi$ to increase at the center of the star, and then to turn around and decrease at larger $r$.  This turnaround can occur because the pressure decreases with $r$, causing the force to increase.

Once the star is large enough that $\phic \approx \phit$, chameleon ``screening'' makes the field far less responsive to further increases in $G\mstar/\rstar$.  Screening is the stellar analog of the chameleon thin shell effect, in which the deviation of $\phi$ from its background value is sourced only by a thin shell of matter near the surface of an object.  The thin shell effect becomes important when the field at the center of an object approaches the minimum of its effective potential inside that object, in a precise analogy to stellar screening.  For a sufficiently large star, the effective stellar mass that acts as a source to $\phi$ will be much smaller than the actual stellar mass.

Specifically, we define the relativistic analogues of the bare (linear) and effective
(screened) masses:
\begin{eqnarray}
\mlin
&=&
\int_0^{\rstar} 4\pi r^2 (\rho-3P) dr,
\\
\mscr
&=&
-\int_0^{\rstar} \frac{3r^2}{2G} \frac{\partial \Veff}{\partial \phi} dr
\nonumber\\
&=&
\int_0^{\rstar}\left[ 4\pi r^2 (\rho-3P) 
  + \frac{3r^2}{2G}\frac{dU}{d\phi}\right] dr.
\end{eqnarray}
In the linear regime, when the slope of the potential is small, $\mscr \approx \mlin$.  Since these are cases where $P/\rho\ll 1$, 
$\mlin  \approx \mstar$  and
the field feels the total mass, which is also the source of the gravitational potential $G\mstar/\rstar$.  
Therefore the change in the field is proportional to the gravitational potential
(see Sec.~\ref{subsec:unscreened_stars} for a more extended treatment).

As the potential and the change in the field become large, the field rolls to a steeper part
of the potential.  Thereafter the field source is screened by the potential, reducing the source from $\mlin$ to $\mscr$.  As $\phic \rightarrow \phit$, this screening becomes complete at the center of the star, and the chameleon is only sensitive to stellar matter at larger $r$.  Fig.~\ref{f:screen} illustrates screening in two different ways.  As the potential $G\mstar/\rstar$ is increased in Fig.~\ref{f:screen}~(left), $\phic-\phids$ increases steadily until $G\mstar/\rstar \approx 1-\phids$.  Beyond that point, $\phi$ is insensitive to further increases in the potential.  Fig.~\ref{f:screen}~(right) shows that the onset of this insensitivity coincides with the decrease of $\mscr/\mlin$.  As $\phic \rightarrow \phit$, the chameleon ``sees'' a smaller and smaller portion of the star, so further increases in the potential are unable to push the field all the way to the curvature singularity.

There has been much confusion in the literature between non-linear chameleon effects and relativistic effects.  As Fig.~\ref{f:screen} makes clear, screening is a chameleon effect that is totally unrelated to strong gravity.  The chameleon enters the non-linear regime when $G\mstar/\rstar \approx \phit-\phids \approx 1-\phids$ and screening becomes important.  This is true even when $1-\phids \ll 1$, for which a star with $G\mstar/\rstar \approx 1-\phids$ is non-relativistic.  In the Starobinsky $f(R)$ model, $1-\phids$ is determined by the choice of model parameters $n$ and $x_1$; for $n=1$, $1-\phids = x_1^{-2}$.  Chameleon effects and relativistic effects will coincide when $x_1$ is of order unity, as in much of the literature.  For $x_1 \gg 1$, chameleon effects will appear in objects with potentials much smaller than unity.   In fact, the appearance
of chameleon effects for the galactic potential is required for
solar system tests of gravity \cite{Hu_Sawicki_2007}.  In other words only Starobinsky models
with $x_{1}\gg 1$ and $1-\phids \lesssim 10^{-6}-10^{-5}$ are in fact viable.

\begin{figure}[tb]
\begin{center}
\includegraphics[angle=270,width=3.5in]{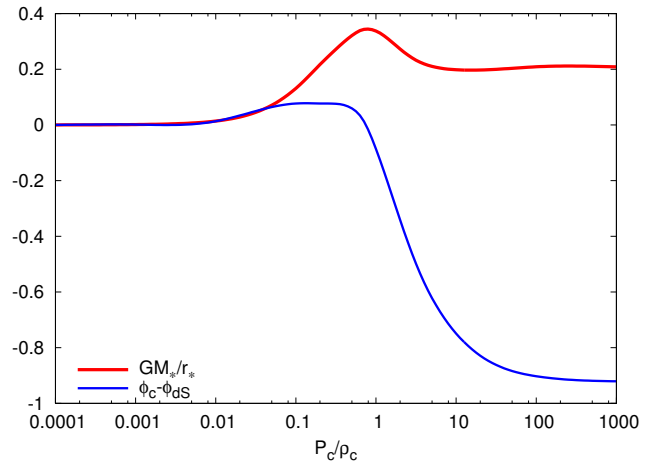}
\caption{Potential and chameleon field in high-pressure stars.  As $\Pc$ is increased beyond $\frac{1}{3}\rhoc$, the potential reaches a maximum and $\phic-\phids$ becomes negative. \label{f:maxphi}}
\end{center}
\end{figure}

Finally, we note that Fig.~\ref{f:screen}~(left) hints at a truly relativistic effect.  For $x_1=3.6$, the plot of $\phic-\phids$ begins to decrease as $G\mstar/\rstar$ is increased beyond about $0.3$, corresponding to $\Pc \gtrsim \rho_c/3$.  Note that general relativity
would require negative scalar curvature $R$ at the center of such a star. 
For stars of this size, $\mlin$ begins to decrease with respect to $\mstar$.  As we continue to increase $\Pc/\rhoc$ beyond $1/3$, we find that $\mlin$ and $\mscr$ can become negative.  This means that $\phic-\phids$ will be negative in a sufficiently high-pressure star, as in the thin curve in Fig.~\ref{f:screen}~(left).  Furthermore, we find that $G\mstar/\rstar$ reaches a peak in the high $\Pc$ regime, and then begins to decrease with $\Pc$, as shown in Fig.~\ref{f:maxphi}.  

The high central pressures $\Pc/\rhoc \gg 1$ do not actually
reflect the average $P/\rhoc$ through the star.  
In fact, for these
stars, the stellar radius, as well as the pressure and field profile near
the surface, become independent of $\Pc/\rhoc$, keeping $G\mstar/\rstar$
constant.
The rapid rise
in the central pressure in a small central core is accompanied by a suppression of the central field $\phic \rightarrow 0$ which allows potential and pressure gradients to
be balanced in hydrostatic 
equilibrium.

For $n=1$ and $x_1=3.6$, we find that the gravitational potential reaches a maximum of $G\mstar/\rstar = 0.345$ at $\Pc/\rhoc = 0.78$.  This is in contrast to general relativity, in which $G\mstar/\rstar$ increases monotonically toward $4/9$ as $\Pc/\rhoc \rightarrow \infty$.  (For the stars considered here, corrections to this general relativistic upper limit due to the presence of a de Sitter horizon are negligible \cite{Andreasson_Bohmer_2009}.)  However, such high pressures are not thought to be realized in any star composed of ordinary matter, so this difference between general relativity and $f(R)$ gravity is unlikely to be useful for observationally distinguishing between these theories.


\section{Analytic Arguments}
\label{sec:screening}

In the previous section, we used numerical examples for several choices of
$f(R)$ parameter values in order to show that:
\begin{enumerate}
        \item the chameleon field $\phi$ can be in one of two regimes, linear and
non-linear;
        \item the linear regime is characterized by a linear scaling between
the field and the gravitational potential, $\phic-\phids = G\mstar/\rstar$;
        \item the non-linear regime is characterized by a field profile that
increases near the center, but turns around before hitting the singularity;
        \item the transition between these regimes occurs when the
gravitational potential approaches $1-\phids$, the distance in field space between the
de Sitter value and the singularity, and is unrelated to relativistic
effects.
\end{enumerate}

Here, we use analytic arguments to generalize these four results to a
broader range of parameter values and stellar densities.  In Sec.~\ref{subsec:unscreened_stars}, the
linear field equation is solved exactly for a non-relativisitic, constant
density star in order to show that $\phic-\phids = G\mstar/\rstar$.  The linear
approximation breaks down rapidly as $G\mstar/\rstar$ approaches $1-\phids$, even in a
non-relativisitic star.  In Sec.~\ref{subsec:screened_stars}, we study the field near the stellar
center in the non-linear regime.  $\phi(r)$ increases slowly in response to the
pressure, even as $-GM(r)/r$ decreases.  The field approaches the singularity,
but turns around before reaching it.  We also use our analytic solution to
show that matching the exterior boundary condition requires exponential precision in setting the central field value.
The implications of this exponential tuning are studied for different
densities and parameter values, such as those used in
\cite{Kobayashi_Maeda_2008a}.


\subsection{Unscreened stars}
\label{subsec:unscreened_stars}

Throughout this section, we work in the non-relativistic limit, $\Pc/\rhoc,\; G\mstar/\rstar \ll 1$, 
in which the equation of motion (\ref{e:sphi}) reduces to
\begin{equation}
\phi'' + \frac{2}{r}\phi' = -\frac{dU}{d\phi} - \frac{8\pi G}{3}\rhoc.
\label{e:sphi_nr}
\end{equation}

Approximating the potential by 
\begin{equation}
\frac{dU}{d\phi} \approx -\mds^2(\phi-\phids),
\label{e:Uapprox}
\end{equation}
valid for $\phi$ near $\phids$, we find 
\begin{equation}
\phi(r)-\phids =
\begin{cases} \frac{8\pi G \rhoc}{3\mds^2} + C_\mathrm{int}\frac{\sinh(\mds r)}{\mds r} 
& r<\rstar ,\\
 C_\mathrm{ext} \frac{e^{-\mds r}}{\mds r}
 & r>\rstar. 
 \end{cases}
\end{equation}

\begin{figure}
\begin{center}
\includegraphics[angle=270,width=3.5in]{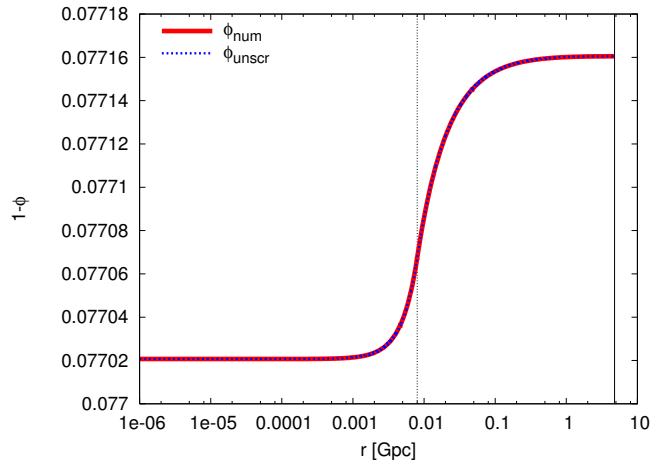}
\caption{Numerical ($\phi_\mathrm{num}$) and approximate ($\phi_\mathrm{unscr}$) field profiles for a star in the unscreened limit.  We assume $n=1$, $x_1=3.6$, $\rhoc=100\rho_\Lambda$, and $\Pc/\rhoc = 10^{-4}$.  The dotted and solid vertical lines correspond to the stellar surface and the horizon, respectively. \label{f:unscreened}}
\end{center}
\end{figure}

The constants $C_\mathrm{int}$ and $C_\mathrm{ext}$ are found by matching $\phi(r)$ and $\phi'(r)$ at the stellar surface, $r=\rstar$, and the resulting field profile is shown in Fig.~\ref{f:unscreened}.
At the stellar center, 
\begin{equation}
\phic-\phids 
= 
\frac{G\mstar}{\rstar}
\left[ \frac{2\left(1 - (1+\mds \rstar)e^{-\mds \rstar}\right)}
  {(\mds \rstar)^2} 
\right].
\end{equation}
The quantity in brackets approaches $1$ in the limit of small $\mds \rstar$, which is an excellent approximation because $\mds \sim H_\Lambda$.  Thus we have $\phic-\phids = G\mstar/\rstar$ for an unscreened non-relativistic star.  Since $G\mstar/\rstar \ll 1$, this
also validates the assumption in Eq.~(\ref{e:Uapprox}).
 


\subsection{Screened stars}
\label{subsec:screened_stars}

As $G \mstar/\rstar$ approaches $1-\phids$, $\phic \rightarrow 1$, so (\ref{e:meff}) implies that $\meff$ becomes large.  Thus, our approximation that $\meff \approx \mds$ is violated; the unscreened scaling breaks, and the field no longer
responds linearly with the potential.  In this screening limit, the full field profile can only be calculated
numerically.
On the other hand, in the stellar interior, where screening is nearly complete, analytic
solutions are available that provide insight into the numerics.

  To zeroth order, the field near the stellar center sits at the minimum of its effective potential, 
\begin{equation}
\frac{\partial}{\partial \phi} \Veff(r,\phimin) 
= 
-\left.\frac{dU}{d\phi}\right|_\phimin -\frac{8\pi G}{3}(\rhoc - 3P) 
=
0.  
\label{e:phimin}
\end{equation}
This is because the field at $r \approx 0$ is insensitive to the field outside the star, which is many Compton wavelengths away, and can adjust itself to minimize the local effective potential.  
Note that $\phimin(0) = \phit$; $\phimin$ is a generalization of $\phit$ to $r>0$.
In particular, we see that $\phimin(r)$ increases with $r$ at the center toward the singularity, because $P(r)$ decreases.  Since the field must eventually match onto the exterior solution, with $\phi'(\rstar)<0$, the field must turn around.  This applies to non-relativistic as well as relativistic stars.  
 Given a fixed pressure profile it is
straightforward to solve for $\phimin$.   In the high curvature, $R \gg R_0$ limit,
the minimum corresponds to the general relativistic expectation that $R=8\pi G (\rho-3P)$, so analytic expressions for $P(r)$ are available.

One can solve the field equation~(\ref{e:sphi}) iteratively to obtain successively
better approximations to the screened solution.   To first order the screened  solution $\phiscr$  becomes
\begin{equation}
-\frac{d U}{d\phi} \Big|^{\phiscr}_{\phimin} = \left[ \phimin'' + 
\left( \frac{2}{r} + \frac{N'}{2N} + \frac{B'}{2B}\right) \phimin' \right] B ,
\end{equation}
where $N$ and $B$ are given by the general relativistic solution.
 
\begin{figure}[tb]
\begin{center}
\includegraphics[angle=270,width=3.5in]{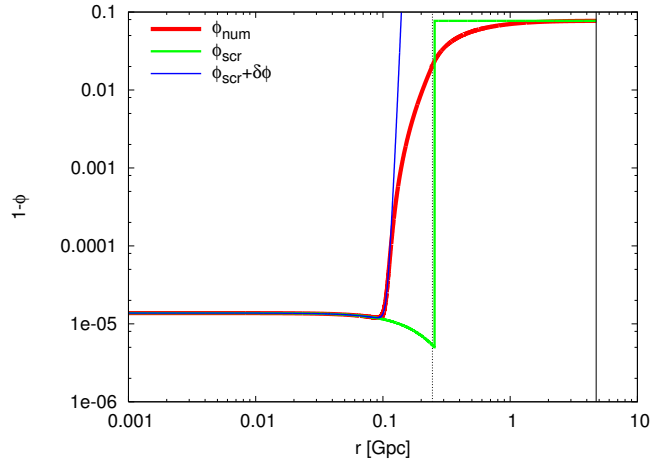}
\caption{$1-\phi$ for the numerical computation, the screening limit $\phiscr$, and  perturbations around it $\phiscr + \delta\phi$ in the constant mass limit of Eq.~(\ref{e:dphi_N0}).  We assume $n=1$, $x_1=3.6$, $\rhoc=100\rho_\Lambda$, and $\Pc/\rhoc = 0.1$.  The dotted and solid vertical lines correspond to the stellar surface and the horizon, respectively.\label{f:phimin}}
\end{center}
\end{figure}

We stop at first order, since
the true solution must depart substantially from the screened solution in the outer regions of the
star in order to match the exterior solution smoothly. 
Fig.~\ref{f:phimin} shows $\phiscr$ compared with the numerical solution of the equations of motion, for a star with $\Pc/\rhoc = 0.1$ in an $f(R)$ model with $n=1$ and $x_1=3.6$.
Notice that the two solutions only deviate in a shell of mass near the stellar radius,
corresponding to a region where the integrand of $\mscr$ becomes nonzero.  Thus the outer regions of the star source the deviation of $\phi$ from $\phiscr$ that allows it to roll continuously to $\phids$. This numerical solution simply reflects a smooth interpolation between the interior screened solution and
the exterior boundary condition.

Deviations from the screened solution at the outskirts imply that even deep in the
 interior 
there will be small deviations.
  These deviations $\delta \phi 
= \phi-\phiscr$ are governed by
\begin{equation}
\left[ \delta \phi'' + 
\left( \frac{2}{r} + \frac{N'}{2N} + \frac{B'}{2B}\right) \delta\phi'  \right] B =
-\frac{d U}{d\phi} \Big|^{\phiscr+\delta\phi}_{\phiscr} 
\label{e:fielddeviations}
\end{equation}
where the $N$ and $B$ solutions can be iterated to the appropriate order.

Note that in the approximation that the field has rolled only a small distance from
its central value, $|\delta\phi| \ll 1-\phit$,
\begin{equation}
-\frac{d U}{d\phi} \Big|^{\phiscr+\delta\phi}_{\phiscr}  \approx m_{\rm t}^2 \delta\phi
\end{equation}
where
\begin{equation}
m_{\rm t}^2
= 
\frac{1}{3}\left.\left(\frac{\phi}{d\phi/dR}-R\right)\right|_{\phi=\phit} \,.
\end{equation}
Equation~(\ref{e:fielddeviations}) therefore becomes 
 source free and has Yukawa-like solutions which
exponentially grow with $\mt r$.   Since $\mt \rstar \gg 1$, this implies that 
$\phi_c$ must be exponentially close to, but not exactly equal, $\phiscr$ at $r=0$, 
in order for the deviations from the screening solution to become significant
only near the stellar radius.  Thus numerical solutions for stars that are screened
for a substantial part of their interior are difficult to find numerically by shooting
from a central value $\phi_c$.

These considerations can be made more concrete for non-relativisitic stars.
In this case, the pressure profile $P(r) \approx \Pc(1-r^2/\rstar^2)$, and
the corresponding zeroth order solution $\phimin$ contains a quadratic piece.
The screened solution deviates from $\phimin$ as
\begin{eqnarray}
m_{\rm t}^2 (\phiscr-\phimin)
& =&  
\phimin'' + \frac{2}{r}\phimin' \nonumber 
\approx 
3\phimin'' \\
&\equiv & m_{\rm t}^2\kappa \approx \frac{(8\pi G)^2 \rhoc^2}{2m_{\rm t}^2} \,.
\end{eqnarray}
Furthermore, deviations away from the screened solution grow as 
\begin{eqnarray}
\delta\phi(r) 
=
\delta\phi(0)
\frac{\sinh(\mt r)}{\mt r} .
\label{e:dphi_N0}
\end{eqnarray}

There are a number of interesting properties of the numerical solution that can
be gleaned from this analytic treatment.
We have already shown that $\phit = \phimin(0)$ is the threshold between field solutions that decrease monotonically ($\phic<\phit$) and those that increase at low $r$ ($\phic>\phit$).  Now we see from Eq.~(\ref{e:dphi_N0}) that, if $\phic > \phiscr(0) = \phit + \kappa $, then the field will monotonically increase.  Thus there is an interval of width $\kappa$ in field space for which the field will increase at the center and then turn around.  For a star with $n=1$, $x_1=3.6$, $\rhoc=100\rho_\Lambda$, and $\Pc=0.1\rhoc$, the width of this interval is of order $10^{-9}$.  Fig.~\ref{f:phimin} also shows $\phiscr + \delta\phi$ for this case 
under the approximation of Eq.~(\ref{e:dphi_N0}), with $\delta\phi(0)$ chosen so as to match the turnaround point from the numerical solution.  
Note that deviations from this approximation are expected {\it after} the field turns around
to match the exterior boundary since the constant mass approximation breaks down.
We also drop the small relativistic correction here for simplicity.

Furthermore, most of this interval $\phit < \phic < \phit + \kappa$ corresponds to fields that turn around at $r \sim \mt^{-1}$, much earlier than in a typical screened star.  If we want this turnaround position $\rturn$, at which $\phi'(\rturn)=0$, to be much larger, then much more tuning in the field value is necessary.  For $\rturn > r_0$, the lower bound on $\delta\phi(0)$ becomes
\begin{eqnarray}
\delta\phi(0)
&>& 
-\frac{2}{3} \kappa
(\mt r_0)^2
e^{-\mt r_0}.
\label{e:dphi0bnd}
\end{eqnarray}
We see from our numerical solution Fig.~\ref{f:phi}~(right) that $\rturn \approx \rstar/2$ for $\Pc/\rhoc=0.1$.  
From (\ref{e:dphi0bnd}), the interval in field space for which $\rturn>\rstar/2$ is of order $10^{-39}$, approximately the amount of tuning that we needed earlier in order to find a numerical solution.  We can estimate the numerical difficulty of finding $\phic$ though the shooting method, for a general star, by expressing $\kappa$ in terms of the $f(R)$ model parameters $n$ and $x_1$,
\begin{eqnarray}
\kappa
&=&
9 \cdot 2^{8n+3} (2n+1)^2
\left[
1-\frac{1+(n+1)x_1^2}{(1+x_1^2)^{n+1}}
\right]^{-2}
\nonumber\\
&&\times
\left(\frac{\rhoc}{\rho_\Lambda}\right)^{-(4n+2)}
x_1^{-4n}.
\end{eqnarray}
At $\rhoc = 2\times 10^6 \rho_\Lambda$, a star with the same $\mt\rstar$ as the one discussed above, and $\rturn > \rstar/2$, will require that $\phic$ be tuned to a precision $10^{-65}$, and this tuning only becomes worse at higher $n$.


\section{Conclusions}
\label{sec:conclusion}

We have studied static, spherically symmetric stars in $f(R)$ theories of modified gravity.  Such theories behave like ordinary gravity with a chameleon field, a matter-coupled scalar with non-linear self interactions.  We have found the chameleon field profile inside a star by numerically integrating the modified Einstein equation.  The critical features of these
numerical solutions are exposed by analytic arguments.

We find solutions to the equations of motion over a wide range of central pressures, $10^{-5} < \Pc/\rhoc < 1000$, and gravitational potentials, $1.4\times 10^{-5} < G\mstar/\rstar < 0.345$ in 
Starobinsky's model of $f(R)$ gravity.   This range of potentials extends from small, sun-like stars all the way to highly relativistic neutron stars.  Thus the existence of
relativistic stars cannot be used to rule out $f(R)$ theories of gravity.

We have shown that, in the non-linear regime of the chameleon field, stars are screened.  That is, $\phi$ is sourced only by the outer portion of a star's mass, analogous to the chameleon thin shell effect.  This keeps $\phi$ from reaching the singularity $\phi_\mathrm{sing}=1$ as $G\mstar/\rstar$ is increased.  As the stellar density increases 
at fixed $G\mstar/\rstar$, the disparity between the Compton wavelength of the field
and the stellar radius  
makes numerical solutions unfeasible, but does not imply that solutions do not exist.

Screening sets in when $G\mstar/\rstar \approx 1-\phids$, that is, when the gravitational potential approaches the distance in field space between the singularity and the de Sitter background.  This marks the transition between the linear and non-linear regimes of the chameleon.  This transition is unrelated to strong gravity.  Its appearance at
$G\mstar/\rstar \approx 10^{-1}$ in KM \cite{Kobayashi_Maeda_2008a} is an artifact of
their choice of an $f(R)$ model with $1-\phids \approx 10^{-1}$. 
We have explored models where  $1-\phids \approx 10^{-3}$ and confirmed that screening sets in at $G\mstar/\rstar \approx 10^{-3}$, that is, in non-relativistic stars.

The fact that chameleon effects can appear at much lower potentials than strong gravity is essential to the construction of  viable $f(R)$ theories \cite{Hu_Sawicki_2007}.  Viability requires that the Galaxy, with potential $\Phi \sim 10^{-6}-10^{-5}$,  be in the non-linear chameleon regime.  It is only in this regime that deviations from general relativity in the solar system
are sufficiently suppressed to satisfy local tests.

\vspace{0.6cm}

{\it Acknowledgments:}
We thank Maria Beltr\'an, Ignacy Sawicki, Fabian Schmidt, Tristan Smith, and Bob Wald for useful conversations.   
 This work was supported by the Kavli Institute for Cosmological
 Physics (KICP) at the University of Chicago through grants NSF PHY-0114422 and
 NSF PHY-0551142.  WH was additionally supported by U.S.~Dept.\ of Energy contract DE-FG02-90ER-40560 and
 the David and Lucile Packard Foundation.
\bibliographystyle{arxiv_physrev}
\bibliography{fRstar}

\end{document}